\documentclass[prodmode,acmtap]{ci2014}

\acmVolume{2}
\acmNumber{3}
\acmArticle{1}
\articleSeq{1}
\acmYear{2010}
\acmMonth{5}

\usepackage[ruled]{algorithm2e}
\SetAlFnt{\algofont}
\SetAlCapFnt{\algofont}
\SetAlCapNameFnt{\algofont}
\SetAlCapHSkip{0pt}
\IncMargin{-\parindent}

\title{YASCA: A collective intelligence approach for community detection in complex networks}
\author{Rushed Kanawati \affil{University of paris Sorbonne Cit\'e }}

\begin{abstract}
In this paper we present an original approach for community detection in complex networks. The approach belongs to the family of seed-centric algorithms. However, instead of expanding communities around selected seeds as most of existing approaches do,  we explore here applying an ensemble clustering approach to different network partitions derived from ego-centered communities computed for each selected seed. Ego-centered communities are themselves computed applying a recently proposed ensemble ranking based approach that allow to efficiently combine various local modularities used to guide a greedy optimisation process.  Results of first experiments on real world networks for which a ground truth decomposition into communities are known, argue for the validity of our approach.
\end{abstract}
\begin{document}

\maketitle

\section{Introduction}
Complex networks are frequently used for modeling interactions in real-world systems in diverse areas, such as sociology, biology, information spreading and exchanging, scientometrics and many other different areas. One key topological feature of real-world complex networks is that nodes are arranged  in tightly knit groups that are loosely connected one to each other. Such groups are called {\em communities}. Nodes composing a community are generally admitted to share common proprieties and/or be involved in  a same function and/or having a same role. Hence, unfolding the community structure of a network could give us much insights about the overall structure a complex network.

We distinguish between two different problems: partitioning the whole graph into (eventually overleaping) communities \cite{For10a} and identifying ego-centered communities for a given query node \cite{A3:KanawatiAFGG2014}. In this work we propose a new algorithm, YASCA, that use local community identification in order to compute a global graph partition into communities. The algorithm belongs to the seed centric algorithms family \cite{A3:KanawatiSCSM2014}.  The basic idea of seed centric algorithms is to select a set of nodes (i.e. seeds) around which communities are constructed.  Being based on local computations, these approaches are very attractive to deal with large-scale and/or dynamic networks.  Different algorithms apply different policies for seed selection and for community construction around seeds. The number of seed nodes can be pre-determined  \cite{KHO10a} or computed by the approach itself \cite{A3:KanawatiSocialcom2011}. The seed selection process can be : random \cite{KHO10a} or informed \cite{A3:KanawatiSocialcom2011}. The community construction can be made applying consensus techniques, expansion techniques  or agglomeration techniques \cite{A3:KanawatiSocialcom2011}.  Next we propose an original seed centric approach that apply an ensemble clustering approach \cite{Strehl2003} to different network partitions derived from ego-centered communities computed for each selected seed.  
 
 \section{YASCA: the proposed algorithm}
 The YASCA\footnote{Yet Another Seed-centric Community detection Algorithm} algorithm is structured into three main steps: 
 \begin{enumerate}
 \item First, we select a subset of nodes acting as seed nodes. Let $S \subset V$ denotes the set of selected seed nodes. Different selection strategies can be applied as mentioned in previous section. A detailed discussion of selection strategies can also be found in \cite{A3:KanawatiSCSM2014}
 \item For each selected node, we compute an ego-centered community using a recently proposed ensemble ranking based greedy optimisation algorithm described in \cite{A3:KanawatiAFGG2014}.  This ensemble ranking based approach allows combining efficiently different local modularities generally used for identifying local communities \cite{ChenZG09}. Let $C_{v}$  be the computed local community of seed node $ v \in V$ . The set of vertices $V$ can then be partitioned into two disjoint  sets : $P_v= \{C_{v}, \overline{C_{v} \}}$ where $\overline{C_{v} }$ denotes the complement of set $C_v$. 
 \item We merge the set of obtained partitions $P_v, v \in S$ by applying a cluster ensemble method. The output of this process is the taken to be the final decomposition of the shoe graph into communities. 
 \end{enumerate}
 
 The goal of an ensemble clustering approach is to compute a clustering (here a partition) that combine the different obtained partitions. One widely applied method is based on constructing a {\bf consensus graph} out of the set of partitions to be combined \cite{FernB04,Strehl2003}.  The consensus graph $G_{cons}$ is defined over the same set of nodes of the initial graph $G$. Two nodes $v_i, v_j \in V$ are linked in $G_{cons}$ if there is at least one partition $P_{Q_x}^y$ where both nodes are in a same cluster. Each link $(v_i, v_j)$  is weighted by the frequency of  instances that nodes $v_i, v_j$ are placed in the same cluster.  Different approaches has been proposed to detect communities in the consensus graph \cite{Strehl2003,DahlinS13}. In this work we propose applying to the consensus graph a community detection algorithm that can handle unconnected, weighted graphs. The Louvain algorithm \cite{Blondel}  is  one good option. 

 \section{Experiments}
\begin{figure}[h]
\begin{center}
\includegraphics[width=0.4\textwidth]{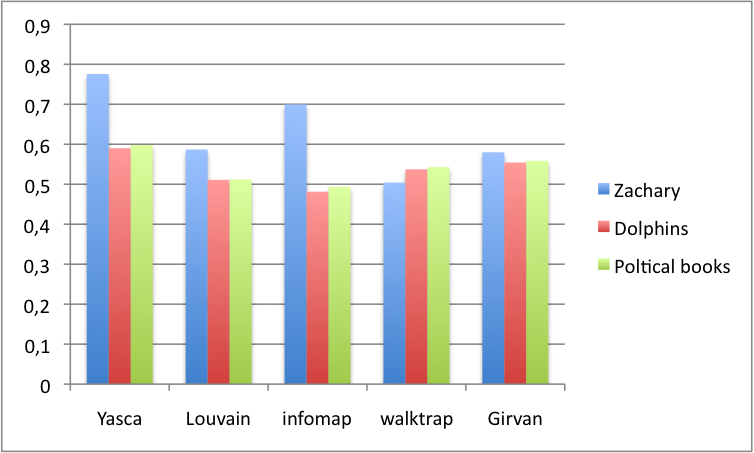}
\label{res}
\caption{Comparative results on the three selected dataset in terms of NMI}\end{center}
\end{figure}

 In order to quantitatively analyse and compare performances of the different proposed approaches we have applied these to networks whose community structure is already known. Performance of the proposed algorithms are evaluated in function of the similarity of the obtained clustering with the ground truth known clustering as measured by the normalized mutual information (NMI) index \cite{Strehl2003}. We compare the performances of the YASCA algorithm on three well known benchmark networks: the Zachary Karate Club network \cite{ZAC77a}, the dolphins social network  \cite{Lusseau2003}, and the political books network  \cite{Girvan2002}. We have configured YASCA as follows : Seed nodes are composed of the 25\% of top high degree nodes and 25\% of low degree nodes. For the consensus graph we keep a link is the associated frequency if equal of greater than 0.5 (these are the best parameters when using the degree centrality for selecting seeds). Figure \ref{res} shows the obtained results on the three datasets compared to state of the art algorithms : Louvain \cite{Blondel}, Infomap \cite{Rosvall09}, Walktrap \cite{PON06a} and edge-betweenness based modularity optimisation algorithm \cite{Girvan2002}. 

 \section{Conclusion}
A new seed centric algorithm for community detection is proposed. First results on small networks show the potential of this algorithm compared to the state of the art algorithms. Further investigations about the effects of different parameters of the algorithm are considered (seed selection strategy, the local community algorithm to be used, etc.). Validations on large -scale graphs are also scheduled. This requires to parallelise the step of local community identification of each seed node.
\bibliographystyle{ci-format}
\bibliography{rk,community,local-community,overlapping-com,rankaggregation,clustering,vote}

\end{document}